\documentclass[compsoc,conference]{IEEEtran}
\IEEEoverridecommandlockouts
\usepackage{cite}
\usepackage{amsmath,amssymb,amsfonts}
\usepackage{algorithmic}
\usepackage{graphicx}
\usepackage{textcomp}
\usepackage{xcolor}
\def\BibTeX{{\rm B\kern-.05em{\sc i\kern-.025em b}\kern-.08em
    T\kern-.1667em\lower.7ex\hbox{E}\kern-.125emX}}

\begin{document}

\title{Machine Learning for Detection and Mitigation of Web Vulnerabilities and Web Attacks\\
}

\author{\IEEEauthorblockN{
Mahnoor Shahid}
\IEEEauthorblockA{\textit{Universität des Saarlandes} \\
\textit{Saarland, Saarbrücken, Germany}\\
mash00001@stud.uni-saarland.de}
}

\maketitle

\begin{abstract}
Detection and mitigation of critical web vulnerabilities and attacks like cross-site scripting (XSS), and cross-site request forgery (CSRF) have been a great concern in the field of web security \cite{top_attacks}. Such web attacks are evolving and becoming more challenging to detect. Several ideas from different perspectives have been put forth that can be used to improve the performance of detecting these web vulnerabilities and preventing the attacks from happening \cite{survey}. Machine learning techniques have lately been used by researchers to defend against XSS and CSRF \cite{Chen_survey}, \cite{Liu_survey}, \cite{Kaur_survey}, and given the positive findings, it can be concluded that it is a promising research direction. The objective of this paper is to briefly report on the research works that have been published in this direction of applying classical and advanced machine learning to identify and prevent XSS and CSRF. 
The purpose of providing this survey is to address different machine learning approaches that have been implemented, understand the key takeaway of every research, discuss their positive impact and the downsides that persists, so that it can help the researchers to determine the best direction to develop new approaches for their own research and to encourage researchers to focus towards the intersection between web security and machine learning.

\end{abstract}

\begin{IEEEkeywords}
Web Security; Web Vulnerabilities; Machine Learning; XSS attacks; CSRF attacks
\end{IEEEkeywords}

\section{Introduction}
Web applications have become a staple of our daily lives due to their widespread use in fields such as banking, finance, marketing, taxation, social interaction, and even medical. 
Many crucial operations are carried out online on the web that includes data which could be personal details like identification information, passwords, pin or pass codes, medical details or financial details like transactions, purchase details, tax filings etc. Although, web applications do provide security and safety measures for protection, there exists some web vulnerabilities which are loopholes, bugs or voids that are exploited, by an adversary, either on client side or server side. Indeed, the number of reported web vulnerabilities \cite{stats1},\cite{stats2} and the frequency of web based attacks \cite{stats3} have been significantly rising each day, necessitating the critical need for stronger security. However, web applications, are extremely complicated and challenging to analyze because of the various complex programming techniques that are used to develop them. Therefore, identifying those web vulnerabilities is not a straight forward task and is one of the critical component of a web application's defense.


Cross-site Scripting (XSS) and cross-site request forgery (CSRF) are considered to be among the top severe risks to web security from the kinds of vulnerabilities demonstrated by web applications, as per OWASP Top Ten (Open Web Application Security Project) \cite{top_attacks}. Many detection algorithms based on conventional techniques, have been provided in order to track down these web vulnerabilities and defend against these attacks \cite{survey}. Unfortunately, they are not sufficiently effective enough against a threat that can take any form and is evolving as time passes by. So, another critical component of a web application's defense is to detect mischievous and carefully crafted web attacks by determining whether or not a target script is malicious.

This is where machine learning comes to the rescue. Machine learning (as well as deep learning) has been widely acknowledged and adopted in many fields to simplify and automate any highly complicated task, improve its performance, and increase scalability. Researchers of web security community have also tried to integrate different machine learning techniques to effectively identify XSS or CSRF attacks \cite{Chen_survey}, \cite{Liu_survey}, \cite{Kaur_survey} and it has proved to be very beneficial. This assertion has been validated by the development and evaluation of many different detection methodologies that has leveraged machine learning (or deep learning) and have been comprehensively covered in this paper. 

In this paper, various research works are reported and discussed that have been published in this direction to inspire other researchers to step in this direction and adopt machine learning or deep learning techniques and use them to resolve web application threats. The underlining contributions of this paper can be summarized as follows:
\begin{itemize}
    \item Provide a concise and systematic literature review of the existing works that have used machine learning and deep learning procedures to detect and mitigate XSS web attacks and vulnerabilities.
    \item Provide a brief overview of the existing works for automating CSRF detection and the advent of utilizing machine learning for classifying sensitive requests.
\end{itemize}

\section{Cross Site Scripting}
Cross-site scripting (XSS) attack\cite{XSS} is a web attack that is sometimes known as a code-injection attack in which the attacker inserts malicious code script, generally JavaScript, around benign code into a legitimate web application using a weakness (XSS vulnerability) in the web server or in the user's browser in order to access cookies, user sessions, and other sensitive data. This attack can also be exploited to execute other web attacks, like CSRF, or DOS attack. The Open Source Foundation for Application Security (OWASP)\cite{OWASP} positions cross-site scripting (XSS) attacks higher on its list of critical web vulnerabilities. \\

\noindent
\textbf{XSS: attack or vulnerability?}
XSS vulnerabilities are security bugs, or flaws in web application that enables the injection of any irregular and harmful (probably JavaScript) code around innocuous code whereas in XSS attacks, an adversary takes the advantage and exploits those XSS vulnerabilities in order to execute the attack by successfully transmitting the malicious code.

\subsection{Classical machine learning approaches for XSS attacks classification}
Initially, researchers have applied classical machine learning strategies to detect and mitigate XSS that usually involved a feature extraction process that focuses on specific target attributes to identify XSS attacks. 
Multiple classifiers are trained on top of those features using various tactics. Then, those models are evaluated to observe the impact of the generated features. The performance of the classifiers are compared and the scores of the best model are used to conclude the generalization capability of the proposed procedure. Discussed below are some interesting research articles that have been published under this umbrella:\newline 

\noindent
\textbf{Obfuscated malicious JavaScript detection using classification techniques [2009]} \\
Likarish et al.\cite{Likarish} developed a machine learning strategy to identify obfuscated malicious JavaScript code snippets. Malicious JavaScript that has been obfuscated can be used to outwit anti-malware software applications or manual examination from being detected as malicious and can be inserted in web pages to spread malware or carry out other unwanted operations. The primary goal of this study is to identify malicious JavaScript that has been obfuscated which may be used to carry out XSS attacks. 
In order to train the models, they have constructed a dataset which consisted of JavaScript scripts that belonged to both benign and malicious class. Next, they have focused on extracting different features manually and settled with a total of 65 features. In the study, features that are extracted are primarily based on obfuscation methods such encoding, encryption, text/string manipulation, and eval functions. They have further explored how significantly those features affect the classification of these malicious scripts. They observed that malicious JavaScript usually has less proportion of tokens as compared to benign JavaScript, and their use of JavaScript keywords also varied significantly. Lastly, they have trained multiple classifiers: SVM,  ADTree, Naive Bayes, and RIPPER for the detection of obfuscation scripts. Ten cycles of 10-fold cross-validation were used to evaluate the classifiers so they can assess how much the performance of every classifier varies statistically. The evaluation results showed that SVM acquired the best precision score=0.92 while JRIP classifier acquired the best recall=0.787 and F2-score=0.806.  \newline 
\textbf{Pros} - 
A novel feature extraction technique has been proposed which was based on obfuscation methods through which the features set was obtained which was sufficient to distinguish between JavaScript which is malicious and JavaScript which is benign. Four distinct classifiers were evaluated in the research, and found that they are quite accurate. The study offers a number of useful real world applications for the classifiers, including filtering or examining dubious JavaScript.
\newline
\textbf{Cons} - Collected dataset was highly imbalanced with 60,000 benign scripts and only 62 malicious scripts. The test set also consisted of extremely less obfuscated scripts which is not a good representation of real-world JavaScript. It also relies on a single source for malicious scripts. Authors have not clearly defined what constitutes obfuscation or maliciousness in JavaScript, which can affect the validity and reliability of their labels and results. 
They have also not discussed the limitations or challenges of their approach, such as the trade-off between detection accuracy and performance overhead, or the ethical implications of suppressing potentially malicious JavaScript. Another drawback, as addressed by the authors, is that the classifier would also classify some benign scripts (for e.g., \textit{Packed JavaScript}) as malicious causing more false positives.  \newline

\noindent
\textbf{Classification of malicious web code by machine learning [2011]} \\
Komiya et al\cite{Komiya} adapted machine learning algorithms to identify and categorize user input that contains malicious web code such as SQL injection attacks and XSS attacks. 
In their research, they have proposed two feature extraction strategies: blank separation (clear partition), and tokenization. The premise of the first strategy was that a document is made up of several terms/phrases that are separated by spaces. Therefore, terms were extracted by splitting the input using blank spaces and the frequency count of every term was used for the calculation of feature weightage. The second strategy was predicated on the notion that malicious web code includes particular tokens that describe its characteristics. Those terms were extracted from the input using the specified set of tokens and the frequency count of every token term was again used to calculate feature weights. Afterwards, they have trained and evaluated SVM, Naive-Bayes, and KNN for detecting exploitable user inputs. They have also evaluated different types of kernel functions in SVM like Linear, Polynomial, and Gaussian. The highest accuracy, precision as well as recall, was achieved by using these feature extraction techniques along with SVM and Gaussian Kernel.
\newline 
\textbf{Pros} - Proposed feature extraction techniques were very effective and it is reflected in the consistency between the accuracy scores of all the classifiers.
Comparison and evaluation was done using three different machine learning methods on a large and diverse dataset of web codes. \\
\textbf{Cons} - Note that, an inaccurate feature weight might result from phrases or terms being separated in a malicious script by characters other than white spaces. This approach may not be effective against new or unknown types of malicious web code that were not in the dataset. Training and testing the machine learning models on big datasets could take a lot of computing time and resources.\newline

\noindent
\textbf{Defending Malicious Script Attacks Using Machine Learning Classifiers [2017]} \\
Khan et al.\cite{Khan} have focused on a real-time, lightweight system with negligible runtime overheads to defend the end user from being attacked by any injurious JavaScript code. In the proposed approach, an interceptor was installed on the web browser of the user as a plugin. Whenever a user sends any web request to the web server to look up for the requested webpage, the server in return generates a response which was directed towards the interceptor which extracts features from web page's source code and feeds it to the trained ML classifier for the classification process. If there was any presence of malicious JavaScript script, the webpage would be banned prior to the web browser interpreting it. Authors have considered training multiple binary classifiers (SVM, J48, KNN and Naive Bayes,) on a subset of relevant features in order to acquire high accuracy and these classifiers were independently tested and the one with the best performance was selected to be used in the interceptor. According the to the findings reported in the paper, three sets of experiments were conducted, keeping some variations in the training procedures and it was observed that the KNN classifier was able to acquire 90.70\% precision rate and demonstrated better results in all the experiments in identifying XSS, compared to the other classifiers.\\
\textbf{Pros} - Proposed solution is lightweight in design, with little runtime overheads. It is platform and browser agnostic.\\ 
\textbf{Cons} - Collected dataset was comprised of very few malicious samples. Precision can be further improved.\\ 

\noindent
\textbf{Detecting Cross-Site Scripting Attacks Using Machine Learning [2018]} \\
Meerani et al\cite{Mereani_Detect} have also considered utilizing machine learning methods to identify XSS attacks by analyzing HTTP requests and preferred to train SVM, KNN and Random Forests for XSS attacks detection. This was the first time Random Forest was considered for XSS classification. They have proposed a way to come up with a feature set that combines the program syntax and program behavioral features. They settled with a total of 59 features. They have investigated and explored how such features have contributed and have led the classifiers to acquire higher accuracy on real world dataset. They have compiled a collection of dataset from multiple sources that has balanced scripts and gives both injurious and benign scripts enough coverage. After conducting the experiments on the test set, it was concluded that the classifiers were successfully able to distinguish between the malicious and benign scripts with high precision and accuracy. Among the trained classifiers, it was observed that, with a precision of a 96.79\%, k-NN outperformed SVM and Random Forest by a little margin.\\
\textbf{Pros} - Extracted feature set were discriminative enough and had a significant impact/influence over the classifiers' performance. This can be validated by the consistent scores of all the trained classifiers. Moreover, models were trained and evaluated on a very large and diverse dataset, consisting of many malicious samples.\\
\textbf{Cons} - It's possible that the input will just be plain text and not even be a script. First-round testing revealed that simple text was not always appropriately categorized.\\

\noindent
\textbf{Preventing Cross-Site Scripting Attacks by Combining Classifiers [2019]} \\
Meerani et al\cite{Mereani_Prevent} have extended their previous work to further improve the performance XSS detection task by combining the classifiers in two ways. First, by stacking the models together using the ensemble technique and second, by building a two stage classifier using the cascading technique. In this research, they have considered five different classifiers, including a neural network to perform the XSS classification. A stacked classifier is developed using these five classifiers as base because such a classifier gives more accurate and robust predictions. A decision tree is also trained from classifying if the input is a plain text or a script. And using both the classifiers together, another cascading classifier is developed to verify user input into a web application. The suggested methodology has two phases. In the first phase, a decision tree evaluates whether the input is a script or plain text. Results of this first classifier (samples that are classified as scripts) are utilized in the subsequent phase where multiple base classifiers (SVM, KNN, Neural Networks, and Random Forest) were trained to predict whether the script is malicious or not. To train the decision tree, dataset consisting of normal and scripts were collected. Text type features were obtained from them which includes attributes like, letters, numbers, spaces, punctuation etc. To train the classifiers of second phase, another labelled dataset was collected, with the script features (same as their previous work, but with some additional improvement) and using the output of these classifier, the meta classifier on top of it is trained. \\
\textbf{Pros} - Improved feature set along with the stacking of base classifiers, yielded results that were more accurate as compared to their previous work. Cascading catered the problem of plain text not being correctly classified. \\
\textbf{Cons} - Still there were some malicious scripts that were misclassified because of base64 encryption that employs letters more frequently as compared to number or symbols. \\
\textbf{Comparision with the previous study} - With the exception of precision, all of the metrics favors the method proposed in the 2019 study as compared to 2018 study. This implies that while the method in the 2019 study can detect more XSS attacks and prevent more false negatives than the approach of 2018 study, it also causes more false positives. \\

\noindent
\textbf{The Detecting Cross-Site Scripting (XSS) Using Machine Learning Methods [2020]} \\
Kascheev et al.\cite{Kascheev} presented an XSS attacks detection model by analyzing HTTP requests and responses using machine learning. In this work, authors have constructed a dataset of 240,000 JavaScript samples that was comprised of both malicious (40,000) and benign (200,000) samples. In the pre-processing stage, data was transformed in the form queries with respective parameters. They have deciphered the entire dataset into ascii feature vectors and used regular expressions to extract the parameters. They have discarded the queries that had a lot of parameters. Then, a dictionary of approximately 3000 tokens was put together. The feature vectors were created by transforming the dataset using the word2vec technique. For training the models, they have considered the classifiers: SVM, Naive Bayes, Logistic Regression, and Decision tree. They have conducted the experiments to assess the performance of their suggested approach. They have split the whole dataset into two splits train set (consisting of 70\% of the whole dataset) and test set (remaining 30\% of the whole dataset). They have trained the model using the trainset and evaluated the performance of the classifiers on the test set. The accuracy acquired by Decision tree is 98.81\%, Naive Bayes Classifier is 65.27\%, Logistic Regression is 83.03\%, and is 71.37\%.\newline
\textbf{Pros} – A sizable and varied dataset of XSS payloads is used to compare various machine learning methods and to evaluate their effectiveness. It shows that the suggested approach can identify XSS attacks with good accuracy and high recall rates. Another key advantage of this approach, as stated by the authors, is that a single classifier can consider all the data at once and also contribute to the production of new data.\\
\textbf{Cons} - There was severe class imbalance. However, sample weighting was performed in the pre-processing phases, and it may have helped to balance the classes. Disconsistency between the accuracy of the classifiers reflects that the feature generation process should be improved. \\

\subsection{Classical machine learning approaches for XSS vulnerability classification}
Researchers have also applied classical machine
learning strategies to detect XSS vulnerabilities or web pages that are vulnerable to XSS attacks. Followings are some of the works under this category: \\

\noindent
\textbf{Automatic classification of cross-site scripting in web pages using document-based and URL-based features [2012]} \\
Nunan et al.\cite{Nunan} have extended the approach of Likarish et al\cite{Likarish} and have focused on XSS vulnerability classification, as it aims to automatically classify web pages that are vulnerable to XSS attacks using machine learning techniques by extracting and analyzing predictive set of features that are based on URL-based and Document-based attributes. The have compiled a dataset of 216,054 samples. Out of which, 15,366 samples were malicious and were obtained from XSSed dataset whereas, benign samples were obtained from Dmoz and ClueWeb09. They have extracted the features and categorized them into three groups: obfuscation based (same as Likarish et al\cite{Likarish}), suspicious patterns, and HTML/JavaScript schemes They have selected two machine learning classifiers, that is SVM and Naïve Bayes. After training those classifiers, the have evaluated the performance of the classifiers under three aspects: rate of false alarms, accuracy or precision, and detection. With a 98 percent accuracy rate for XSS classification, SVM classifier had the best overall performance, although Naive Bayes classifier also had satisfactory performance that was comparable to SVM.\\
\textbf{Pros} - Eight alternative machine learning algorithms are compared in the study with this technique, demonstrating the reliability and universality of its features. Accuracy and performance have increased with the enhanced set of features as validated by a quick comparison of performances with the features mentioned by Likarish et al.\cite{Likarish}. \\
\textbf{Cons} - This technique may be quite accurate, however it only uses one source for malicious scripts. The research does not address the computational complexity or scalability of its methodology, which may have an impact on its applicability and effectiveness in real-world circumstances.
\newline

\noindent
\textbf{Machine Learning Based Cross-Site Scripting Detection in Online Social Network [2014]} \\
Wang R. et al.\cite{Wang} developed a machine learning strategy to identify XSS in social networks. They have extracted features related to online social network, from the web pages and grouped them into four categories: HTML tag features, JavaScript features, keyword features, and URL features They have built a database that consisted of 29,046 benign web pages from DMOZ, 13,935 XSS vulnerable webpages from XSSed as well as 8,063 webpages that can simulate XSS worms (weibo.com). They have compiled a feature database, consisting of features that encapsulates the attributes mentioned before. To extract these features automatically, they have developed a feature extractor that takes the collected web pages as its input and generates the corresponding feature vectors as its output. 
Using this feature database, they have trained two classifiers: an alternative decision tree (ADTree) and an adaptive boosting (Adaboost). They have evaluated the performance of their approach, and it was reported that the higher precision and recall were obtained with AdaBoost algorithm, the scores were precision=0.941, recall=0.939 and F1 score=0.939. Although the scores for ADTree were also comparable with precision=0.938, recall=0.936 and F1 score=0.936. \\
\textbf{Pros} - Suggested features have shown to be effective. The caliber of the extracted features has a direct impact on the classification effect. \\
\textbf{Cons} - XSS vulnerable webpages and XSS worms vulnerable pages were from a single source. \\
\newline
\noindent
\textbf{Predicting Cross-Site Scripting (XSS) security vulnerabilities in web applications [2015]} \\
K Gupta et al. \cite{Gupta} proposed a method to use predictive models based on text mining to find XSS susceptible code files in web applications. Their proposed technique is based on using a customized way of tokenizing text features (combining basic and context features). This method converts each code into a specific collection of text features with a corresponding frequency. They have settled with 20 features list (combining basic and context features). Using the WEKA tool, they have conducted all the experiments and have repeated all the experiments 10 times with randomly chosen train and test splits. According to their reported evaluation results, bagging classifier outperformed the other classifiers (SVM, Naive Bayes, Random Forest, etc.) and was able to achieve the performance accuracy of 92.6\% with the F score of 90.6\% \\
\textbf{Pros} - The argument made in the study is that current XSS vulnerability prediction techniques do not take into account the user input context in the output statement, which is essential for discovering context-sensitive vulnerabilities. The performance of the  model was greatly enhanced by taking into account the user's input's context. Moreover, the study assesses the effectiveness of several machine learning classifiers using a dataset of 1000 PHP files gathered from a number of open source projects. \\
\textbf{Cons} - Effectiveness of this methodology is constrained with detecting malicious code from the PHP application's source files.
\newline

\noindent
\textbf{Detection of XSS in web applications using machine learning classifiers [2020]} \\
Banerjee, et al.\cite{Banerjee} utilized machine learning algorithms to extract attributes that might be helpful to find vulnerabilities that XSS poses to web applications. They have focused on generating the features which are based on JavaScript and URL based attributes. They have constructed a dataset of total 1,611 JavaScript samples with 24 attributes out of which 1,453 samples were benign whereas only 158 were malicious. They have assessed the accuracy and the performance of the extracted features by using four different classifiers: Random Forest, KNN, Logistic Regression and SVM. To validate the performance of their approach they have further evaluated the trained models on test set. Random Forest acquired the highest accuracy of around 98\% for the XSS attack classification. \\
\textbf{Pros} - Projected approach was very straightforward yet effective. This approach utilizes both JavaScript and URL features to detect various XSS variants and evaluates the approach using a diverse dataset. \\
\textbf{Cons} - Machine learning for XSS detection has certain significant challenges that are not addressed, including scalability, robustness, flexibility, and generalization which needs to be further investigated. It would have been better if the feature engineering process targeted towards more complex and generalizable features.

\subsection{Limitations of Machine Learning approaches}

Mentioned below are some major limitations of using machine learning approaches for XSS classification:
\begin{itemize}
    \item The accuracy performance of a machine learning classifier is highly reliant on the manually designed features that are provided to it.  
    \item Unfortunately, the generalization capability of machine learning algorithms is also limited to the previous patterns and relationships in data.
    \item Since XSS attacks may take many different forms and is more likely to change frequently. Feature extraction can be very difficult in order to capture unseen patterns. Therefore, fast evolution of XSS patterns may outpace the performance of conventional machine learning techniques.
    \item Not suitable for complex and large datasets.
    \item Machine learning algorithms cannot compete against the performance of deep learning methodologies.
\end{itemize}

\subsection{Deep learning for XSS attacks classification} 
Due to the fact that shallow models strongly rely on the previous patterns and reasoning, malicious behaviors that deviates distinctively in feature space from those seen before fail to be classified, leading to low accuracy scores and less scalability. In contrast, deep neural networks are more robust and achieve better generalization. Without the need for extensive feature engineering, deep models are capable of autonomously learning high-level invariant representations from the training dataset and are able to accurately classify even the samples that were not exposed while training and deviate reasonably. Thus, it is fairly evident in \cite{deep_learning} that deep learning approaches are exceptionally better than classical machine learning approaches in terms of detection accuracy, false alarm rate, and scalability.
Discussed below are some research publications that have applied robust and trending deep learning architectures to detect and mitigate XSS attacks: \newline 

\noindent
\textbf{A Deep Learning Approach for Detecting Malicious JavaScript Code [2016]}\\
Wang et al. \cite{Wang_dl} have applied deep learning techniques to classify malicious JavaScript code accurately. They have proposed a deep learning framework utilizing the stacked denoising autoencoder model that automatically learns the deep features for pattern recognition. They have formed a dataset of 27,103 samples in total out of which 12,320 belong to the benign class and 14783 belongs to the malicious class. Samples in the dataset were either obfuscated or plain JavaScript. In the preprocessing stage, every input was translated into binary feature vectors by converting every character in the code sample into 8bit binary codes and all the segments were saved as a binary file. By taking the advantage of sparse random projections, authors have reduced the dimensions of the dataset up to 480 dimensions in order to lower the computational cost and time. After they have trained the SdA network, they evaluated the performance of their proposed approach SdA-LR and from the reported results, it can be concluded that it was able to outperform the previous machine learning algorithms with lower inference time and better detection accuracy. \\
\textbf{Pros} - SdA was able to optimize the performance accuracy of Logistic Regression and SVM. 
Inference time was comparatively very low. No extensive feature engineering was required. It can deal with JavaScript code that is obscured, which is a typical tactic employed by adversaries to avoid detection.\\
\textbf{Cons} - False positive rate of the proposed approach was around 4.2\%, which is relatively very high. 
Training time was considerably very long. Required more computing resources. Requires an extensive amount of labeled data for training.\\

\noindent
\textbf{DeepXSS: Cross Site Scripting Detection Based on Deep Learning [2018]}\\
Fang Y et al. \cite{Fang} presented a deep learning-based strategy for XSS attacks detection called “DeepXSS”. They have first preprocessed the input by passing it via a decoder that iteratively employs every possible decoding schemes. The next step in the preprocessing process was generalization of the decoded data where URLs were substituted with “http://website”, numbers with “0”, strings with “param\_string” and special characters were eliminated. Last step of preprocessing was tokenization of the input. After that, they have used word2vec to extract the features which encapsulates the semantic knowledge of the XSS payloads. Then, they have trained a RNN based architecture, LSTM (long short term memory) network using those feature embeddings. 33,426 Malicious samples were gathered from XSSed database whereas 31,407 benign samples were gathered from the DMOZ database for conducting the experiments. From the experimental results, as reported by the authors, we can assess that the LSTM based detection method can be very effective, since it was able to reach a precision of 0.995 and a recall of 0.979. \\
\textbf{Pros} - This research was a valuable nudge towards this specific direction of applying LSTMs (or deep learning methodologies) for XSS detection. Dataset used for experiments was also nearly balanced.\\
\textbf{Cons} - Focus on complete coverage of all the security-sensitive aspects of XSS was not the part of this research. Moreover, using embeddings can make the training process exceedingly time-consuming, and cumbersome.
\newline

\noindent
\textbf{MLPXSS: An Integrated XSS-Based Attack Detection Scheme in Web Applications Using Multilayer Perceptron Technique [2019]}\\
F. M. M. Mokbal et al. \cite{Mokbal_2019} have presented a rudimentary but a novel ANN based model for identifying XSS attacks and have integrated that with an adaptable feature extraction procedure that extracts digital dataset as features, based on URL, HTML, and JavaScript based attributes of the datasets.
For their research, authors have constructed a huge uniformly randomized dataset, comprised of 138,569 distinctive records with benign samples (100,000) and malicious samples (38,569) of 41-dimensions. 
Later, a feed forward network (multi layer perceptron model) is trained for the classification of XSS attacks. This approach was assessed on the unseen testset of 27,714 samples (with 19,967 benign and 7,747 malicious samples) to prove the viability of the suggested model and acquired an overall accuracy of 99.32\% with precision=0.99, and recall=0.98. After comparatively evaluating the proposed method with the existing methods (since the malicious samples were gathered from XXSed), it was concluded that the proposed approach relatively better. \\
\textbf{Pros} - Feature extractor technique was able to obtain suitable and more relevant feature vectors that was able to differentiate the XSS anomaly more precisely. By utilizing a low complexity deep neural network model in conjunction with a flexible and adaptable feature extractor, makes it platform independent. Both the client-side and the server-side of a system can use it as a security layer 
\\
\textbf{Cons} - Limitations of this research includes the real-time detection of XSS attacks.
Detection rate of 98.35\% can also be improved. \\

\noindent
\textbf{XSS Detection Technology Based on LSTM-Attention [2020]}\\
Lei L et al. \cite{Lei} proposed another LSTM based model that additionally utilizes the attention mechanism for distinguishing XSS. At first, by taking the advantage of decoding procedure at the preprocessing phase, they have eliminated the encodings from the datasets. Next, from the datasets, they have omitted the HTML and URL encodings, to decrease the dimensions of data. They have further standardized the dataset by performing some replacements like, substituting a number with “0” and URLs with “http://u”. In order to extract the features from the datasets, they have developed a number of unique regular expressions, and then utilized word2vec (for vectorization) to create feature vectors for training the network. Utilizing an LSTM network, abstract features of an XSS attack were obtained and attention mechanism was applied to enhance the classification performance. Proposed LSTM-Attention model was later evaluated and after conducting, a number of experiments, it was apparent that the model's classification performance is improved by adding the attention method as it was able to acquire significantly better precision of 99.3\% and recall of 98.2\% in successfully classifying XSS. \\
\textbf{Pros} - Precision score, recall score, and F1 score all escalated, when attention mechanism was incorporated into the LSTM network, by 2.9\%, 1.9\%, and 2.1\%, respectively. \\
\textbf{Cons} - Generalization accuracy of the model can be further enhanced, if other types of XSS data is gathered in order to capture more complex and varying form of XSS.

\noindent
\textbf{Cross Site Scripting Attacks Classification using Convolutional Neural Network [2022]}\\
Kumar J et al.\cite{Kumar} have developed a CNN based architecture for XSS classification. They have constructed a dataset from XSSed that consisted of 17,750 samples which includes both malicious and benign scripts. The obfuscation from the scripts was first eliminated. They have translated all the characters of the XSS code scripts into their ASCII representations (reshaped and resized into [120,120]), eliminating non-ASCII characters. To standardize the feature matrix, it was further divided by 128. They transformed the vector into a [120,120] two-dimensional matrix, which they have used as an input for CNN network. Their proposed CNN technique was able to acquire the accuracy of about 98\%. \\
\textbf{Pros} - General characteristics of a CNN architecture does not require extensive pre-processing feature extraction as compared to other classification ML algorithms. The strategy which authors have used for feature development, takes into account every character in the dataset, as well as their order and sequence. As a result, the proposed method is more adapted for continuously evolving XSS. \\
\\
\noindent
\textbf{Detect Cross-Site Scripting Attacks Using Average Word Embedding and Support Vector Machine [2022]}\\
F. M. M. Mokbal et al.\cite{Mokbal_2022} have proposed NLP-SVM approach for detecting web-based XSS threats. NLP was used to pre-process the XSS attack payloads and the classification task was performed by employing a SVM model. These payloads were transformed into vectors at the payload level. For each attack payload, a vector was obtained at payload-level for each sample by averaging the word-level. From the complete dataset, a total of 20,257 vectors of 96 dimensions each were extracted. Using a robust, supervised machine learning technique, such as SVM, the generated vectors were then utilized to model. A thorough analysis of the reported results revealed that the suggested approach could easily and effectively identify XSS-based attacks with exceptionally higher accuracy along with lower false negative (0.4\%) and false positive (1.0\%) rates.\\
\textbf{Pros} - The proposed approach successfully outperformed the well-known 8 algorithms that made use of the identical data by a wide margin. With a low percentage of FP (false positives) and FN (false negatives), the proposed method achieved higher accuracy and an incredible detection rate.\\
\textbf{Cons} - Attack payloads could be automatically extracted from the content using some sort of automated method. \\

\subsection{Deep Learning for XSS vulnerabilities classification}
Researchers have also applied deep
learning strategies to detect XSS vulnerabilities. Followings are
some of the works under this category:\\

\noindent
\textbf{Link: Black-Box Detection of Cross-Site Scripting Vulnerabilities Using Reinforcement Learning [2022]}\\
Lee S et al.\cite{Lee} have aimed at automatically detecting the reflected XSS vulnerabilities by utilizing the reinforcement learning in a black box setting. \\
\textbf{Background} - Although, there are several penetration tools to detect this form of XSS attack which are easy to use and deploy, unfortunately there are some drawbacks to that. Penetration tools uses a predefined attack payload dictionary and it does not consider the context of the target payload that is context-unaware payload. If the dictionary does not contain a working payload then the detection will fail, leading to false negatives. Moreover, if the dictionary is considerably large and the exploitable payload exist in the end of the dictionary, there will be a lot of unnecessary requests and it will increase request overhead of the penetration testing.  \\
\textbf{Proposed Solution} - In this research, authors have presented “Link” which leverages reinforcement learning to generate context-aware payloads. Link generates XSS attack payloads using generation and mutation rules; after executing each payload, Link takes feedback from a target application to automatically generate context-aware payloads.\\
\textbf{Reinforcement Learning} - The agent and environment, both are required for reinforcement learning. The agent's role is to decide on an action based on the state from the given environment and then the chosen action will modify both the environment and the environment's state. Next, the environment computes reward to determine if the selected action is effective. After giving the agent the incentive and the state information, the agent decides the next action based on the reward and state. That is how the agent learns the best course of actions to maximize the overall reward, by repeating these steps. This iterative procedure is what authors have used to identify cross-site scripting vulnerabilities, where the action corresponds to a generation or mutation rule for producing a payload, the environment corresponds to a target application and the reward and state corresponds to a feedback from the target application. \\
\textbf{Objectives} - reducing the number of requests and increasing true positives were two primary goals of this research.
\textbf{Implementation} – The first design component is “action”. Link supports 2 action types: generation and mutation. We have seven generation rule that generates the basic payload for further mutation and we have 32 mutation rules that mutate the basic payloads from the generation.
The next design component is “state”, which provides the information of the environment to the agent. It contains 47 features and these feature vectors consist of scaler values that represents the respective feature’s existence or its type. These features can be divided into three types: information about the input payload, the injection point and about the attack result. The first type of the state is the payload information and these features tell information about the appearance of the payload. The second type of the state is the injection point information. The injection point represents whether the attack payload is placed in the response. The last type of state is information about the attack result. From the attack response, Link analyzes how the injected payload is reflected. 
The last design component is “reward”. So, it gives a positive reward when the attack succeeds and gives bigger reward as the number of attempted requests is lesser. Also, Link gives the negative reward, if the agent chooses the same actions or tries the same payload because it is it is inefficient to try a failed attempt again. \\
\textbf{Evaluation} – After conducting extensive experiments, Link was evaluated against 4 state-of-the-art penetration testing tools and the reported results proved that Link outperformed them by detecting the highest number of bugs with lower number of requests.\\
\textbf{Pros} – In contrast to earlier methods, Link does not require manual intervention to assess selected behaviors in given environments. Link can detect vulnerabilities in the shortest possible time. The benchmark was not included in the training set yet Link still managed to acquire highest score in vulnerability detection. The prospective effectiveness of utilizing reinforcement learning to discover reflected XSS vulnerabilities is further shown by Link, that also uncovered 43 vulnerabilities in 12 real-world apps, which none of the scanners used in the evaluation were able to discover \\
\textbf{Cons} -  The amount of time needed for training and testing increased by 20\% when utilizing a headless browser.\\

\section{Cross Site Request Forgery}
Cross-site request forgery, commonly known as CSRF \cite{csrf}, is one of the most simplest yet effective web attack. CSRF is a web-vulnerability in which any malicious website instructs or coerces a victim’s web browser to send a request to a target website that is vulnerable to perform any security sensitive operation and these malicious requests appear to be a victim's interactions with the target website, taking advantage of the victim's network access and browser settings like cookies.

\subsection{Static and rule-based CSRF defenses}
Followings are some popular mitigations that exist to prevent CSRF attack\cite{defenses_csrf}:
\begin{itemize}
\item	Use CAPTCHAs, one-time passwords, or enforce re-authentication to block cross-site requests from passing through undetected. 
\item	Authenticating a private/secret request token.
\item	Verifying the HTTP request headers.
\item	Verifying custom headers attached to XMLHTTPRequests
\end{itemize}

But unfortunately, each of these mechanisms has their own pitfalls and are not entirely satisfactory since they may not work in some cases. In order to maintain a balance between security and usability, need for automated CSRF approaches was comprehended. \\

\noindent 
\textbf{Automatic Black-Box Detection of Resistance Against CSRF Vulnerabilities in Web Applications} \\
Another black-box, language agnostic CSRF detection method was proposed method by Sadeghi et al.\cite{Sadeghi} that passively identifies CSRF-resistant requests by accessing and examining the browser-server traffic to the target website which does not causes unauthorized database changes to compare the application behavior to a deliberately fabricated request. In this research, authors have defined some set of rules to investigate the presence of anti-CSRF tokens in traffic and determines whether a request is potentially vulnerable or not. \\

\noindent 
\textbf{Same Site Cookies comes to the rescue} \\
Same Site cookies\cite{same_site} is another recently proposed and a very effective mechanism, that is intended to provide defense against the whole family of cross site attacks. Same site cookies are technically generic cookies but with a supplementary attribute of ‘same site’ which is controlled and regulated by the website that sets the cookie with the labels ‘lax’ or ‘strict’. If the value of the same site attribute is set to 'strict' then the cookie is most likely to be excluded from any cross site requests. But in the case of attribute value ‘lax’, the cookie might be included in top-level cross-site GET requests. Same site cookies are supported by popular browsers like Opera, Firefox\cite{firefox}, Chrome\cite{chrome}, and Edge\cite{Microsoft} and browsers that doesn’t support same site cookies, simply handle them as regular cookies.

Multiple blogs and comprehensive studies \cite{same_site_study1}, \cite{same_site_study2} have been published that have thoroughly evaluated the efficacy of same site cookies. 

Applicability, adaption, effectiveness, and in-depth evaluation of the same site cookie was thoroughly explored and investigated at a large-scale in the study by S. Khodayari and G. Pellegrino\cite{same_site_study2}, where they have extensively conducted experimentations to assess it’s practical usage as well as it’s impact; that is if it is set to default, it can cause disruption in functionalities. Authors have also looked over to the threats against the efficiency of same site cookies and the security coverage provided by the Lax policy.

Authors have systematically revealed different weak areas and issues of the same site cookie policy which is a beneficial step to improve by fixing it's problems but correspondingly they have also stated the limitations and the need to transcend above static, pre-packaged policies by discovering more adaptable and customizable same site policies. Such methods require careful and accurate use of security measures, to capture full protection without negatively affecting the user experience. 


\subsection{Towards automated CSRF detection}
When it comes to XSS vulnerability detection, there has been a lot of efforts that has been invested to put forward many approaches but most of the prior works presented to detect CSRF vulnerability detection is heavily inclined towards manual tactics and are mainly white-box.

Therefore, an automated approach for preventing CSRF was first introduced through Deemon\cite{Deemon} which is a model-based framework for security testing intended to detect CSRF vulnerabilities with dynamic analysis, using program status changes, data flow patterns, and graph properties and is built on the PHP interpreter's runtime monitor. 

\subsection{Employing ML procedures for automated and black box CSRF detection}
Even though Deemon has proven to be effective there are certain shortcomings of this approach. Firstly, it is not a language-agnostic analyzer since it only operates on PHP applications and secondly, it requires the accessibility of the source code for dynamic analysis. \\

\noindent 
\textbf{Mitch: A Machine Learning Approach to the Black-Box Detection of CSRF Vulnerabilities [2019]} \\
Calzavara et al.\cite{Mitch} offered an automated black-box solution Mitch to overcome these limitations by utilizing ML approach, for the first time, to classify HTTP request as "sensitive" or "non-sensitive" automatically. Authors have approached this task in a supervised manner and for that they had to come up with a dataset of 5828 HTTP requests (out of which 939 were sensitive requests), which they constructed by accumulating requests from 60 popular websites (according to Alexa ranking) and they have published it for future researches. 

Subsequently, they have done feature engineering, as they were not only interested in classifying sensitive HTTP requests precisely accurately but also interested to know which features were interpretable and led to certain predictions. This is also the primary 

They have further used their constructed dataset to train and test multiple binary classifiers Logistic Regression, Decision Trees, Random Forests, SVM, and Gradient Boosted Decision Trees and have selected the classifier which contributed the best ROC AUC scores which was Random Forest with a score of 0.932 on training set and 0.924 on test set (along with a F1 score of 0.72). They have also assessed the performance of the best classifier with the existing methods CsFire\cite{Csfire} and BEAP\cite{beap}.

Using the best-performing classifier as a building block, authors have developed a browser extension called Mitch; which is a language-agnostic tool to identify CSRF, which functions without getting the access to the web application's source code.

Lastly, they have done some experiments where they have concluded the accuracy of Mitch which achieved a score of 0.78 F1 score on 20 top Alexa websites and they have also compared it’s performance with Deemon on 3 PHP applications where it exceled. They have concluded that there aren't many FP (false positives) or FN (false negatives) returned by Mitch. \newline 

\noindent
\textbf{Incompetent aspects of the suggested technique}
In this research, authors have focused on certain type requests that they have considered to be “sensitive” and those are “visibly sensitive requests” that changes any security-relevant web application state which is visible at the browser side when processing the corresponding response. Keeping this definition in mind, HTTP requests were collected and labelled as sensitive. Since using this strategy binary classifiers were trained, it is certainly irrelevant to compare the performance of CsFire\cite{Csfire} and BEAP\cite{beap} because they carries a different definition of what they consider to be sensitive. Also, Mitch will not be able to cater all the server side changes that has no direct impact on the browser side.

Although broader coverage is undoubtedly a crucial feature of real-world security. However, authors have not focused on complete coverage of all the security-sensitive features and have only conducted a much brief navigation session in all the trials because the prototype does not have any crawler component. 

Researchers have simulated CSRF attacks using Mitch on sensitive requests, which may lead to inappropriate changes to the application database due to its invasive nature.

Because there was a severe class imbalance in the dataset, they have gone for stratified random sampling to make sure there is even distribution of sensitive and non-sensitive request in training and testing. Conversely, it would have been interesting, if rather than training binary classifiers, authors have resolved this task using one class classification.

\section{Burp Suite meets Machine Learning}

Burp Suite is a Java-based framework that offers an extensive set of tools for web and mobile app penetration testing. It accomplishes this by depicting and controlling all traffic between the attacker and a web server. Burp suite is considered as a state-of-the-art scanner for web vulnerabilities. In this proposed work\cite{burp}, researchers have applied advanced ML algorithms to the Burp Suite extension for detecting vulnerabilities like Cross-Site Request Forgery(CSRF), SQL injection, and XML External Entity.

\section{Related Works}

Many surveys have been released that reviewed different methods to detect and prevent XSS or CSRF web attacks. However, only a few number of publications, have been published under the general heading of using various machine learning approaches to identify and counteract CSRF and XSS attacks.

Since Calzavara et al. \cite{Mitch} were the first authors to apply machine learning to tackle a CSRF attack, there has been no study so far that has focused on covering automated or utilizing machine learning to cater CSRF attacks. 

For the purpose of classifying XSS, Chen, X. et al.\cite{Chen_survey} described many approaches that have applied different machine learning techniques XSS web detection. Their reporting was very brief and they didn't include any noteworthy deep learning-based approaches.

An extensive survey of XSS detection methods was issued by M. Liu e al.\cite{Liu_survey}, but unfortunately their focus was not completely on targeting machine learning approaches to cater XSS attacks. However, they covered a few machine learning based methods.

In 2021, Kaur J. et al\cite{Kaur_survey} published a survey that exclusively aimed at the works of applying machine learning for XSS detection and mitigation. Unfortunately, they have focused on 10-11 recent works, between 2018 to 2021 and only reported two deep learning based approaches. Many of the significant works were not included in their coverage.

According to my knowledge, this is the first survey to date, that has included most of the research works that have been published in both the machine learning and deep learning settings.

\section{Conclusion}
In this paper, several research studies are discussed that have tackled web attacks like XSS and CSRF, using conventional machine learning and advanced deep learning methodologies. After going through different approaches, it can be concluded that, although the focus of all research works was to prevent the attacks, but every research is limited to certain constraints. Therefore, there is no single, generic method that can be applied to protect the web application from all the variants of a certain attack, entirely. However, to come up with a better and near-to-perfect mechanism it is important to have a better understanding of the attack, it’s characteristics under different aspects, and it’s impact on the web application. Researchers shouldn’t try to accommodate simple attacks that has less impact and rather should try to go for more complex and difficult attacks that has major impact, in such a way, that can reduce the rate of false alarms as much as possible, increase the accuracy so that no attack is missed, generic enough to be deployed for different variants without much tuning, scalable enough to cater attacks that are evolving with time. 

This survey can help researchers to understand how differently XSS or CSRF have been resolved using the power of machine learning. So far, we can establish from the remarkable results in the conducted experiments that it is a compelling direction to achieve accuracy, scalability, and generalization. This can also be observed that how challenging this simple task can be and there has been methods, mentioned in this study, that has provided a way to tackle the attacks, but have some shortcomings which can be further investigated. And if there has been a study that has only focused on a certain variant of attack but one feels promising enough and can further take it to see how it works for other variants, can also be very beneficial.

Reviewing several research papers and publications on XSS detection led to the deduction that a substantial portion of research has been done on XSS classification using machine learning and deep learning, whilst reinforcement learning has not been given much attention up until the submission of Link by Lei L. et al.\cite{Lee}, that has changed the whole perspective of how else researchers can resolve the problem of XSS web vulnerabilities. 

Regarding CSRF threats, Deemon\cite{Deemon} and Mitch\cite{Mitch} were two notable initiatives that considerably reduced the challenge of classifying CSRF vulnerabilities automatically. Nevertheless, if the concerns raised by \cite{same_site_study2} are remedied, same site cookies could be a viable solution.

\end{document}